\documentclass[a4paper,11pt]{article}
\usepackage{pos}
\usepackage{mathtools}
\usepackage{xcolor}
\usepackage{graphicx}
\usepackage{hyperref}
\usepackage{microtype}
\usepackage{cleveref}
\usepackage{booktabs}
\usepackage{xcolor}
\usepackage{tabu}
\usepackage{siunitx}
\usepackage{floatrow}

\newcommand{\twoBaryonNumerics}[0]{Inoue:2011ai,Berkowitz:2015eaa,Wagman:2017tmp,Francis:2018qch,Iritani:2018vfn,Horz:2020zvv,Green:2021qol,Amarasinghe:2021lqa,Lyu:2022tsd}

\newcommand{\luscherPlusExtensions}[0]{Luscher:1986pf,Rummukainen:1995vs,He:2005ey,Kim:2005gf,Lage:2009zv,Bernard:2010fp,Fu:2011xz,Hansen:2012tf,Briceno:2012yi,Guo:2012hv,Briceno:2014oea}

\newcommand{\justExtensions}[0]{Rummukainen:1995vs,He:2005ey,Kim:2005gf,Lage:2009zv,Bernard:2010fp,Fu:2011xz,Hansen:2012tf,Briceno:2012yi,Guo:2012hv,Briceno:2014oea}

\newcommand{\toystudies}[0]{Guo:2012hv,Briceno:2021xlc}

\title{The Lüscher scattering formalism on the $t$-channel cut}

\author*[a]{André Baião Raposo}
\author[a]{Maxwell T. Hansen}

\affiliation[a]{Higgs Centre for Theoretical Physics, School of Physics and Astronomy,\\
The University of Edinburgh, Edinburgh EH9 3FD, UK}

\emailAdd{a.baiao-raposo@sms.ed.ac.uk}
\emailAdd{maxwell.hansen@ed.ac.uk}

\abstract{The Lüscher scattering formalism, the standard approach for relating the discrete finite-volume energy spectrum to two-to-two scattering amplitudes, fails when analytically continued so far below the infinite-volume two-particle threshold that one encounters the $t$-channel cut. This is relevant, especially in baryon-baryon scattering applications, as finite-volume energies can be observed in this below-threshold regime, and it is not clear how to make use of them. In this talk, we present a generalization of the scattering formalism that resolves this issue, allowing one to also constrain scattering amplitudes on the $t$-channel cut.}

\FullConference{%
39th International Symposium on Lattice Field Theory - Lattice2022 \\
8-13 August 2022 \\
Bonn, Germany
}

\begin{document}
\maketitle

\section{Introduction \& motivation}
\label{sec:intro}

In recent years, there has been considerable progress in the determination of two-nucleon and other two-baryon scattering amplitudes using numerical lattice QCD~\cite{\twoBaryonNumerics}. One of the leading methods in these calculations is to first extract the finite-volume energy spectrum and subsequently the scattering amplitudes via the Lüscher formalism and its extensions~\cite{\luscherPlusExtensions}. In such calculations, each finite-volume energy level constrains or predicts the scattering matrix for all multi-hadron channels that can physically propagate at that energy.

One limitation in all finite-volume formalisms to date is that they neglect volume effects associated with $t$-channel (or left-hand) cuts.%
\footnote{For ref.~\cite{Meng:2022oel}, the issue of the cut may be circumvented by working in the plane-wave basis, but this is not specifically discussed in those proceedings.} This is most obviously a problem when the lattice calculation predicts energies that are on top of the cut, as recently seen in ref.~\cite{Green:2021qol}. The finite-volume formalism is manifestly not applicable here, leading to predictions of a real-valued K-matrix (equivalently, a real-valued scattering phase shift) in a region where the latter is known to be complex.

In this proceedings, we present an extension of the original formalism that can be applied on the left-hand cut. We begin with a brief review of infinite-volume scattering and the standard Lüscher formalism, in sections \ref{sec:infvol} and \ref{sec:luscher}, respectively. In section~\ref{sec:problem}, we illustrate how the $t$-channel cut becomes an issue and, in section~\ref{sec:solution}, we briefly describe our approach to a solution, the full details of which will be presented in a publication (to appear). Conclusions and an outlook are given in section~\ref{sec:conclusion}.

\section{Two-to-two scattering in infinite-volume}
\label{sec:infvol}

We review a few properties of scattering amplitudes in the infinite-volume context, making no reference yet to the finite-volume formalism. Considering two-to-two elastic scattering of non-identical mass-degenerate spin-zero particles with physical mass $M$, we write the total four-momentum in a given frame as $P = (E, \boldsymbol P)$ and introduce the standard Mandelstam invariants $s$ and $t$. Mandelstam $s$ satisfies $s = P^2 = E^2 - \boldsymbol P^2 = (E^\star)^2$, where $E^\star$ denotes the centre-of-mass frame energy. Note we will use $\star$ to denote quantities boosted to the centre-of-mass frame.

\begin{figure}
\centering
\includegraphics[width=11.5
cm]{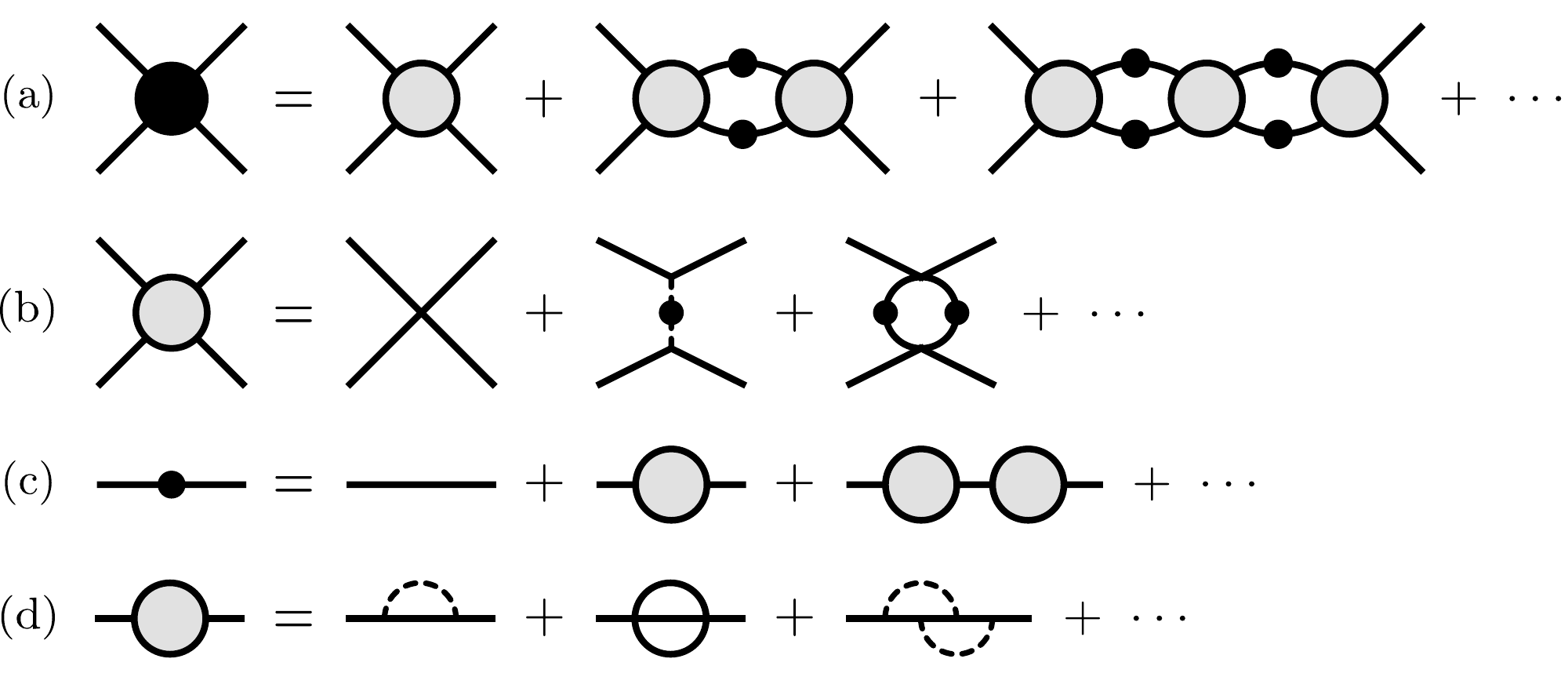}
\caption{(a) Diagrammatic representation of the two-to-two scattering amplitude using Bethe-Salpeter kernels and dressed propagators. (b) Definition of the Bethe-Salpeter kernel as the sum of all connected and amputated two-to-two diagrams which are two-particle irreducible in the $s$-channel. Dashed lines denote other particles that might couple to the scattering channels of interest. (c) Definition of the dressed propagator in terms of bare propagators and self-energy kernels. (d) Definition of the self-energy as the sum of all one-particle irreducible diagrams.
}
\label{fig:bs_kernels}
\end{figure}

The scattering amplitude, which we write $\mathcal{M}(s,t)$, can be formally expressed as the sum of all connected and amputated two-to-two Feynman diagrams, with legs amputated and set on the mass shell (i.e.~with external momenta $p$ having $p^2 = (p^0)^2 - \boldsymbol p^2 = M^2$). This all-orders sum can be organized by introducing the Bethe-Salpeter kernel, defined as the sum of all connected and amputated two-to-two diagrams that are two-particle irreducible in the $s$-channel%
\footnote{In other words, the Bethe-Salpeter kernel is built from diagrams that cannot be separated into two pieces by cutting through two propagators whose momenta sum to the total four-momentum $P = (E, \boldsymbol P)$.}. The amplitude is then expressible in terms of the Bethe-Salpeter kernels and pairs of dressed propagators of the scattering scalars, as shown in figure~\ref{fig:bs_kernels}. Note that all propagators considered are taken with the standard $i\epsilon$ prescription, and all loop momenta are integrated over all components.

It is also instructive to define partial-wave amplitudes according to
\begin{equation}
\label{eq:Mpwaves}
\mathcal{M}(s,t) = \sum_{\ell = 0}^\infty P_\ell(\cos \theta^\star) \mathcal{M}_\ell(s)\, ,
\end{equation}
where $P_\ell$ is a Legendre polynomial and $\theta^\star$ is the scattering angle in the centre-of-mass frame, satisfying $\sin^2(\theta^\star/2) = - t/(s - 4M^2)$.

Using unitarity of the scattering matrix, one can show that the imaginary part of $\mathcal{M}_\ell(s)^{-1}$ is independent of the details of particle interactions. The real part is then typically parameterized using the scattering phase shift $\delta_{\ell}(s)$. We may write
\begin{align}
\text{Im} \, \mathcal{M}_\ell(s)^{-1} = - \rho(s) \, \Theta(E^\star - 2 M ) \, , \qquad
\text{Re} \, \mathcal{M}_\ell(s)^{-1} = \rho(s) \cot \delta_\ell(s) \equiv \mathcal{K}_\ell(s)^{-1} \, ,
\end{align}
where $\rho(s) \equiv \frac{p^\star}{8\pi E^\star} $ is the phase-space factor for non-identical particles, with $p^\star \equiv \frac{1}{2}\sqrt{s - 4M^2}$ denoting each particle's centre-of-mass spatial momentum magnitude, and we have defined the K-matrix $\mathcal{K}_\ell(s)$. This leads to the standard form of the partial-wave amplitude
\begin{equation}
\label{eq:MtoKrelation}
\mathcal{M}_\ell(s) = \frac{1}{\mathcal{K}_\ell(s)^{-1} - i \rho(s)} = \frac{8\pi E^\star}{ p^\star \cot\delta_\ell - ip^\star } \, .
\end{equation}
One can also reach these results via the Bethe-Salpeter series of figure~\ref{fig:bs_kernels} if one defines the K-matrix $\mathcal{K}$ by the same series as the amplitude $\mathcal{M}$, but in which all two-particle loops are evaluated with a principal-value prescription instead of the $i \epsilon$ prescription.

The relations above hold only for $(2M)^2 < s < (E^\star_{\sf inel.})^2$, where $E^\star_{\sf inel.}$ is the lowest-lying inelastic threshold coupling to the channel of interest. This fact has received significant attention for energies above $E^\star_{\sf inel.}$ (in the form of three-particle finite-volume formalisms \cite{Hansen:2014eka,Hansen:2015zga,Hammer:2017kms,Doring:2018xxx,Mai:2017bge,Mai:2018djl}), but here we are concerned with the range $s < (2M)^2$. For these sub-threshold energies, one can analytically continue the amplitude by taking $- i \rho(s) \to \vert \rho(s) \vert$ in order to remain on the physical Riemann sheet. Such an analytic continuation leads to a real-valued scattering amplitude, provided that the K-matrix is real. When, however, we have a lighter particle coupling to the scattering channel of interest, the K-matrix partial waves become complex-valued due to a sub-threshold branch cut, the so-called $t$-channel or left-hand cut. Before turning to the consequences of the cut, we review the standard finite-volume formalism of L{\"u}scher and show that a real-valued K-matrix is implicitly assumed in the sub-threshold analytic continuation, and thus that the formalism is not applicable on the cut.

\section{Review of the L\"{u}scher formalism}
\label{sec:luscher}

In this section, we review the derivation of the L\"{u}scher quantization condition \cite{Luscher:1986pf}, subsequently extended in refs.~\cite{\justExtensions} to include all types of coupled two-particle channels. We focus here on the case of a single channel with two mass-degenerate but non-identical spin-zero particles.

Consider a quantum field theory defined in a finite cubic spatial volume of side-length $L$, with periodic boundary conditions. This system has a discrete $L$-dependent energy spectrum, and the energies lying below the lowest-lying three- or four-particle threshold can be used to extract the elastic two-to-two scattering amplitude. We follow closely the derivation of Kim, Sachrajda, and Sharpe \cite{Kim:2005gf}, used also in refs.~\cite{Hansen:2012tf,Briceno:2015csa,Briceno:2015tza,Briceno:2017max}. We begin by defining a two-point correlation function
\begin{equation}
C_L(E,\boldsymbol P) \equiv \int dx^0 \int_L d^3 \boldsymbol{x}\, e^{-i E x^0} e^{i \boldsymbol{P}\cdot \boldsymbol{x}} \langle 0 \vert \text{T} \mathcal{A}(x) \mathcal{A}^\dagger (0) \vert 0 \rangle_L \, ,
\end{equation}
where the subscript $L$ in $\int_L d^3 \boldsymbol x$ indicates that the integral runs over the finite volume, $E$ denotes the total energy, $\boldsymbol P$ is the total spatial momentum, and $\mathcal{A}(x)$ and $\mathcal{A}^\dagger(x)$ are annihilation and creation operators carrying the quantum numbers of the scattering channel of interest.

\begin{figure}
\centering
\includegraphics[width=14.5cm]{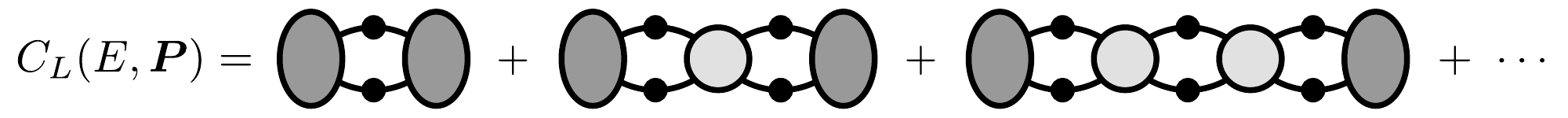}
\caption{Skeleton-expansion representation of the finite-volume correlator $C_L(E, \boldsymbol P)$, in terms of Bethe-Salpeter kernels and dressed propagators, as defined in figure~\ref{fig:bs_kernels}. The end-cap ``blobs'' stand for functions in momentum-space originating from the Fourier transforms of the creation and annihilation operators. As discussed, one can take the kernels and propagators to be the infinite-volume objects, as the difference between these and their finite-volume counterparts is exponentially suppressed, and thus only the loops explicitly shown need to be treated as finite-volume loops.}
\label{fig:fv_corr}
\end{figure}

One can construct a diagrammatic representation for this correlator using the ingredients already introduced in the previous section, the Bethe-Salpeter kernel and dressed propagator pairs. This is known as the skeleton expansion for the correlator and is shown in figure~\ref{fig:fv_corr}. In finite volume, spatial loop momenta are discretized as $\boldsymbol{k} = \frac{2\pi}{L} \boldsymbol{n}$, with $\boldsymbol{n} \in \mathbb{Z}^3$, and we have spatial loop momentum sums instead of integrals, i.e.~we replace $\int \frac{d^3 \boldsymbol k}{(2\pi)^3} \rightarrow \frac{1}{L^3}\sum_{\boldsymbol k \in (2\pi/L) \mathbb{Z}^3}$ for all loops.

The key observation here is that not all loops have the same volume dependence:~loops with intermediate states that cannot go on shell in the energy range considered have exponentially suppressed volume effects $\mathcal{O}(e^{-m L})$ with respect to their infinite-volume analogues, with $m$ being the mass of the lightest particle coupled to the system, while loops with intermediate states that can go on shell have power-like effects $\mathcal{O}(L^{-n})$, for some non-negative integer $n$. In the elastic regime, on-shell states come precisely from the loops left explicit in the skeleton expansion shown in figure~\ref{fig:fv_corr}. Every other loop, implicitly included in the Bethe-Salpeter kernels and the dressed propagators, may be replaced by its infinite-volume counterpart, as the difference, which we neglect, is exponentially suppressed in $L$. Thus, we effectively replace the finite-volume Bethe-Salpeter kernels and dressed propagators with their infinite-volume counterparts.

The contribution from a generic two-particle loop shown in figure~\ref{fig:fv_corr} can be written as
\begin{equation}
C^{\sf loop}_L(P) \equiv \int \frac{dk^0}{2\pi} \frac{1}{L^3} \sum_{\boldsymbol k} \mathcal{L}(P, k) \, \Delta(k) \, \Delta(P - k) \, \mathcal{R}^*(P, k) \, ,
\end{equation}
with $P \equiv (E, \boldsymbol P)$ and loop momentum $k \equiv (k^0, \boldsymbol k)$. The functions $\mathcal{L}$ and $\mathcal{R}^*$ stand for the objects before and after the given loop, $\Delta$ is a fully dressed scalar propagator. Performing the $k^0$-integral and decomposing $\mathcal{L}$ and $\mathcal{R}$ in spherical harmonics with $k$ individually put on-shell, i.e.~setting $k = (\omega({\boldsymbol k}), \boldsymbol k)$ with $\omega(\boldsymbol k) \equiv \sqrt{\boldsymbol k^2 + M^2}$, we can obtain
\begin{equation}
C^{\sf loop}_L(P) =
\frac{1}{L^3}\sum_{\boldsymbol k} \mathcal{L}_{\ell m}(P,\vert \boldsymbol k^\star\vert )\, i \mathcal{S}_{\ell m; \ell' m'}(P, \boldsymbol k; L) \, \mathcal{R}^*_{\ell' m'}(P,\vert \boldsymbol k^\star \vert) + r(P) \, ,
\label{eq:loop_cont}
\end{equation}
where sums over the repeated indices $\ell,m$ and $\ell',m'$ are implied, and we have introduced
\begin{equation}
\mathcal{S}_{\ell m; \ell' m'}(P, \boldsymbol k; L) \equiv \frac{4\pi \, Y_{\ell m}(\hat{\boldsymbol k}{}^\star) \, Y^*_{\ell' m'}(\hat{\boldsymbol k}{}^\star)\, H(\boldsymbol k^\star) }{2 \omega(\boldsymbol k) \, 2 \omega(\boldsymbol P - \boldsymbol k) \, (E - \omega(\boldsymbol k) - \omega(\boldsymbol P - \boldsymbol k) ) } \left(\frac{\vert \boldsymbol k^\star \vert}{p^\star}\right)^{\ell + \ell'}\,,
\label{eq:S_def}
\end{equation}
for later convenience. The term $r(P)$ in eq.~\eqref{eq:loop_cont} is a sum over a smooth summand, leading to exponentially suppressed finite-volume corrections, which we may neglect. The summand in the first term and, more specifically, the quantities $\mathcal{S}_{\ell m; \ell' m'}(P, \boldsymbol k; L)$, contain the pole corresponding to the two-particle intermediate state going on-shell, as can be seen explicitly in eq.~\eqref{eq:S_def}, and thus this term contains all the power-like volume dependence arising from the loop we are considering.

In eq.~\eqref{eq:S_def}, we use $p^\star \equiv \frac{1}{2} \sqrt{s - 4M^2}$, the scattering particle's centre-of-mass spatial momentum magnitude, as defined in section \ref{sec:infvol}. Consequently, setting $\vert \boldsymbol k^\star \vert = p^\star$ satisfies the intermediate two-particle state on-shell condition $E = \omega(\boldsymbol k) + \omega(\boldsymbol P - \boldsymbol k)$. This relation fixes the magnitude of $\boldsymbol k^\star$ at the pole, but not its direction. The barrier factor $\big ( \vert \boldsymbol k^\star \vert/p^\star \big )^{\ell+\ell'} $ is introduced to ensure that no singularities arise from the spherical harmonics.

The function $H(\boldsymbol k^\star)$ is a regulator function which takes a value of 1 for $4 \omega(\boldsymbol k^\star)^2 < (E^\star_{\sf{inel.}})^2$ and of 0 for $4 \omega(\boldsymbol k^\star)^2 > (E^\star_{\sf uv})^2$, where again $E^\star_{\sf{inel.}}$ is the lowest lying three- or four-particle threshold, and $E^\star_{\sf{uv}}$ is some chosen high ultraviolet cut-off. In the region between, $H(\boldsymbol k^\star)$ interpolates smoothly between the two values. This regulator function is similar to the one found in the three-body scattering formalism of refs.~\cite{Hansen:2014eka,Hansen:2015zga}, and corresponds to a separation of low-energy and high-energy parts of the sum.%
\footnote{It should be emphasized that we have renormalization and regularization schemes keeping the overall result finite, the regulator function here simply ensures we have a separation of high-energy and low-energy contributions to the sum such that both parts are finite and that the low-energy part, which contains the relevant singular behaviour, is tractable when implemented numerically. $E_{\sf uv}$ will simply be a scheme-dependence of the formalism, but should be kept high, as setting it too low will lead to enhanced finite-volume effects.}

We next reduce eq.~\eqref{eq:loop_cont} by expanding the functions $\mathcal{L}_{\ell m}(P, \vert \boldsymbol k^\star \vert)$ and $\mathcal{R}^*_{\ell' m'}(P, \vert \boldsymbol k^\star \vert)$ about $\vert \boldsymbol k^\star \vert = p^\star $ and subtracting and adding an integral to reach
\begin{align} \nonumber
C^{\sf loop}_L(P) & = \mathcal{L}_{\ell m}(P, p^\star)\, iF_{\ell m; \ell' m'}(P; L)\, \mathcal{R}^*_{\ell' m'}(P, p^\star) + r'(P) \,, \\
& = {\mathcal{L}}_{\sf os}(P)\, iF(P; L)\, {\mathcal{R}}_{\sf os}^\dagger(P) + r'(P)\, ,
\label{eq:loop_cont_final}
\end{align}
where we introduce the sum-integral difference
\begin{equation}
F_{\ell m;\ell' m'}(P; L) \equiv \left[ \frac{1}{L^3}\sum_{\boldsymbol k} - \ {\sf p.v.} \int \frac{d^3 \boldsymbol k}{(2\pi)^3} \right] \mathcal{S}_{\ell m; \ell' m'}(P, \boldsymbol k; L) \, .
\end{equation}
Here, $\sf{p.v.}$ means the integral is evaluated using a principal-value prescription. The remainder term $r'(P)$ differs from $r(P)$, but still contains the sum of a smooth summand together with $\sf p.v.$ integrals. In the second line of eq.~\eqref{eq:loop_cont_final}, we have defined a compact vector-matrix notation in the angular-momentum index space.

Applying this procedure iteratively to all loops in the skeleton expansion diagrams, it can be shown that we may write the finite-volume correlator in the form:
\begin{align} \nonumber
C_L( P) & = \sum_{n=0}^\infty A(P)\, iF(P; L) \left[i\mathcal{K}(P) \, iF(P; L)\right]^n A^\dagger(P) + C_\infty( P) \\
& = A(P)\, {i} \left[ F^{-1}(P;L) + \mathcal{K}(P) \right]^{-1} A^\dagger(P) + C_\infty (P)\, .
\end{align}
Here, we have summed the geometric series of the first line to obtain the second line. $A(P)$ and $A^\dagger(P)$ are vectors in angular momentum index space, originating from the source and sink operators, and $\mathcal{K}(P)$ is the K-matrix introduced in the previous section%
\footnote{Note $\mathcal{K}$ is an infinite-volume scalar, and thus only depends on $s = P^2$, but we keep $P$ as an argument for compactness.}. Given that we neglect exponentially suppressed volume effects, the $L$-dependence is entirely contained within the matrix $F(P; L)$.

Using a spectral representation of the correlator, it is straightforward to show that it must have poles at the energy levels of the finite-volume spectrum $E_n(\boldsymbol P; L)$. These poles in $C_L(E,\boldsymbol P)$ can only arise from the $L$-dependent part of the first term, meaning that we must have
\begin{equation}
\det \left[ F^{-1}(E_n(\boldsymbol P; L), \boldsymbol P; L) + \mathcal{K}(E_n(\boldsymbol P; L), \boldsymbol P) \right] = 0 \,,
\label{eq:Luescher_cond}
\end{equation}
at all finite-volume energy levels. This is called the \emph{L\"{u}scher quantization condition}, and it can be used to determine $\mathcal{K}$, and hence the scattering amplitude $\mathcal{M}$, from the knowledge of the finite-volume spectrum. The matrices involved in the condition \eqref{eq:Luescher_cond} are formally infinite-dimensional, since the set of possible angular momentum indices $\ell m$ is infinite. For practical use, we must truncate them to the lowest harmonics, making the approximation that $\mathcal{K}$ vanishes for $\ell > \ell_{\sf max}$. This relies on a fast convergence of the partial-wave expansion of the amplitude, such that keeping the lowest harmonics still leads to a reasonable reconstruction of the amplitude.

\section{The \texorpdfstring{$t$}{t}-channel problem}
\label{sec:problem}

The finite-volume spectrum can sometimes include energy levels that drop below the infinite-volume elastic threshold at $s= (2M)^2$. This can occur due to the appearance of a bound state (such that the $L \to \infty$ limit gives the bound-state mass) as well as to an attractive scattering state (such that the energy approaches $2M$ for $L \to \infty$). In many cases, the Lüscher formalism can be analytically continued from $s>(2M)^2$ and the sub-threshold finite-volume energy then provides an important constraint on the K-matrix below threshold.

A subtlety arises, however, when the sub-threshold amplitude $\mathcal{M}(s,t)$, and therefore also the K-matrix, has a nearby $t$-channel cut. This is generically the case in baryon-baryon systems, for example, where a light meson can be exchanged in the $t$-channel.
Taking $m$ and $M$ to be the meson and baryon masses, respectively, and assuming $m \ll M$, one finds that a pole arises in $\mathcal M(s,t)$ at $t = m^2$. For a given fixed choice of the
centre-of-mass frame scattering angle $\theta^\star$, this then leads to a pole in $s$ at
\begin{equation}
s = 4 M^2 - \frac{t}{\sin^2(\theta^\star/2)} \, \bigg\vert_{t = m^2} \, = 4 M^2 - \frac{m^2}{\sin^2(\theta^\star/2)} \ .
\end{equation}

\begin{figure}
\centering
\includegraphics[width=0.95\textwidth]{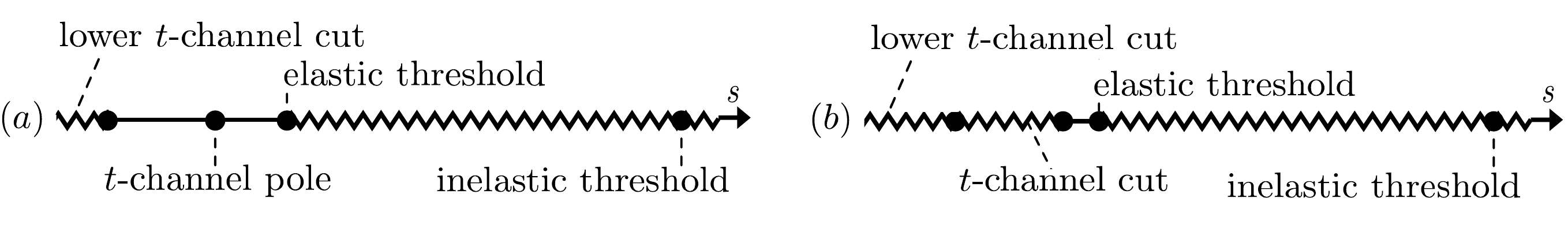}
\caption{(a) Analytic structure of the two-to-two scattering amplitude $\mathcal{M}(s,t)$ in the complex-$s$ plane, for fixed centre-of-mass scattering angle $\theta^\star$, in the case where a lighter particle of mass $m$ couples to the scattering particles of mass $M$. We show the infinite-volume elastic threshold (at $s = (2M)^2$) and inelastic threshold (at $s = (2M + m)^2$) and corresponding branch cuts. Below threshold, we see the $t$-channel exchange pole, corresponding to $t = m^2$, and a lower branch cut, corresponding to two mesons being exchanged in the $t$ channel. (b) Analytic structure of the amplitude when projected to definite angular momentum. The $t$-channel cut, which runs down from the branch point at $s = (2M)^2 - m^2$, arises from the $t$-channel pole.}
\label{fig:an_struc}
\end{figure}

The analytic structure of the scattering amplitude for such systems is shown in figure~\ref{fig:an_struc}(a).
From the expression, one sees that the pole position in $s$ varies from $s = 4 M^2 - m^2$ to $s = - \infty$ as $\theta^\star$ is varied from $0$ to $\pi$. As a result, the angular-momentum projection of the scattering amplitude leads to a branch cut running over this interval as shown in figure~\ref{fig:an_struc}(b).
Multiple meson exchanges can also occur, leading to additional cuts in both the fixed-$\theta^\star$ and the angular-momentum projected amplitudes. In the latter case, these run along $s \leq (2 M)^2 - (n m)^2$ for $n$ exchanged mesons.

As stressed above, finite-volume energies can arise in the region of the branch cuts (as has recently been identified in ref.~\cite{Green:2021qol}) and a naive application of the analytically continued Lüscher formalism fails. In this work, we restrict attention to the region $(2M)^2 - (2m)^2 < s < (2M)^2 - m^2$, in which only the single-meson cut arises, and derive a modified version of the scattering formalism that resolves this limitation.

\section{Proposed solution}
\label{sec:solution}

The breakdown in the original formalism can be traced back to the steps between eq.~\eqref{eq:loop_cont} and \eqref{eq:loop_cont_final} in the review of section~\ref{sec:luscher}. In the step of replacing $\mathcal{L}_{\ell m}(P, \vert \boldsymbol k^\star \vert)$ and $\mathcal{R}^*_{\ell' m'}(P, \vert \boldsymbol k^\star \vert)$ with the on-shell quantities $\mathcal{L}_{\ell m}(P, p^\star)$ and $\mathcal{R}^*_{\ell' m'}(P, p^\star)$, the derivation assumes that the product of the two-particle pole and a given difference, e.g.~$\mathcal{L}_{\ell m}(P, \vert \boldsymbol k^\star \vert) - \mathcal{L}_{\ell m}(P, p^\star)$, is a smooth function of $\boldsymbol k^\star$. This step fails in the sub-threshold region due to the $t$-channel cut.

\begin{figure}
\floatbox[{\capbeside\thisfloatsetup{capbesideposition={right,top},capbesidewidth=9.5cm}}]{figure}[\FBwidth]
{\caption{Separation of the Bethe-Salpeter kernel into a modified kernel (square box), which is safe to put on shell, and the $t$-channel meson exchange, represented by a meson propagator at physical mass.}\label{fig:kern_sep}}
{\includegraphics[width=5.3cm]{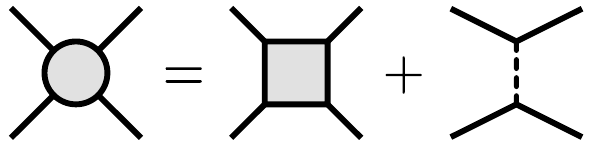}}
\end{figure}

To handle this issue, we separate out the problematic $t$-channel exchanges from the Bethe-Salpeter kernel. We define
\begin{equation}
ig^2T(\boldsymbol k^\star, \boldsymbol k'^\star) \equiv -ig^2\frac{1}{- (\boldsymbol k^\star - \boldsymbol k'^\star)^2 - m^2 + i\epsilon}\, ,
\label{eq:T_def}
\end{equation}
and define a modified kernel by subtracting this from the full Bethe-Salpeter kernel as shown in
figure~\ref{fig:kern_sep}.
Here, $g$ denotes the effective baryon-meson-baryon coupling. We emphasize that $m$ is the physical mass of the meson, and thus that $-iT$ corresponds to the singular part of the fully-dressed meson propagator. The difference between the bare and dressed propagators is smooth, and is simply absorbed into the modified kernel.

Examining the modified kernel, we know that it can be safely evaluated at $\vert \boldsymbol k^\star \vert = p^\star$ and does not possess a singularity or cut in the region $(2M)^2 - (2m)^2 < s < (2M)^2 - m^2$. Crucially, we note also that $ig^2 T$ is safe if kept partially off shell, namely if we keep $\vert \boldsymbol k^\star \vert$ and $\vert \boldsymbol k'^\star \vert$ real.

Recalling expression \eqref{eq:loop_cont_final} for the contribution of a skeleton expansion loop, we again emphasize that the finite-volume frame momentum $\boldsymbol k$, and hence $\boldsymbol k^\star$, are discretized and can be indexed by $\boldsymbol n \in \mathbb{Z}^3$. Thus, we can then treat $\boldsymbol k^\star$ as an extra index, writing $\mathcal{L}_{\boldsymbol k^\star \ell m} \equiv \mathcal{L}_{\ell m}(P,\vert \boldsymbol k^\star\vert ) $, $\mathcal{R}_{\boldsymbol k^\star \ell m} \equiv \mathcal{R}_{\ell m}(P,\vert \boldsymbol k^\star\vert ) $ and defining
$ S_{\boldsymbol k^\star \ell m; \boldsymbol k'^\star \ell' m'}(P; L) \equiv \frac{1}{L^3}\ \delta_{\boldsymbol k^\star \boldsymbol k'^\star} \mathcal{S}_{\ell m; \ell' m'}(P, \boldsymbol k; L)$ such that we may rewrite \eqref{eq:loop_cont_final} as:
\begin{align}
C^{\sf loop}_L(P) & = \mathcal{L}_{\boldsymbol k^\star \ell m}(P)\, iS_{\boldsymbol k^\star \ell m; \boldsymbol k'^\star \ell' m'}(P)\, \mathcal{R}_{\boldsymbol k'^\star \ell' m'}(P; L) + r( P) \,, \\
& = \mathcal{L}(P)\, i S(P; L) \, \mathcal{R}(P) + r( P)\, .
\end{align}
In the first line, we are also implicitly summing over the momentum indices. In the second line, we again employ a compact vector-matrix notation, but now in the angular momentum plus spatial loop momentum index space.

Applying this to all loops in the skeleton expansion diagrams and rearranging by factors of $S(P; L)$, we obtain the finite-volume correlator in the form
\begin{equation}
C_L( P) = \sum_{n=0}^\infty A(P)\, iS(P; L) \left[ ( i\Bar{K} + ig^2 T(P) ) \, iS(P; L) \right]^n A^\dagger(P) + C^{(i)}_\infty( P) \, ,
\end{equation}
where all quantities in the first term are vectors or matrices in the angular momentum plus loop momentum index space. Note that $A(P)$ and $A^\dagger(P)$ are different from those in \eqref{eq:loop_cont_final}. The matrix $i g^2 T$ is the matrix of angular momentum projections of the $t$-channel exchange defined in \eqref{eq:T_def}, and $\bar{K}(P)$ is the sum of all possible smooth contributions one can obtain between $S(P; L)$ matrices. The second term $C^{(i)}_\infty( P)$ is a collection of $L$-independent terms.

From the discussion above, we know it is safe to set $\vert \boldsymbol k^\star \vert = p^\star$ for $\bar{K}(P)$ and, therefore, we can expand $\bar{K}(P)$ about the on-shell point. We implement this by making use of a trivial vector $u$ in the momentum index space, whose elements are $u_{\boldsymbol k^\star} = 1$, and making the substitution $\bar{K}(P) = u \Bar{\mathcal{K}}(P) u^\dagger + \left[\bar{K}(P) - u \Bar{\mathcal{K}}(P) u^\dagger\right]$. The matrix $\bar{\mathcal{K}}(P)$ is a matrix in the angular momentum index space only and corresponds to $\bar{K}(P)$ with the dependence on the magnitude of spatial momentum (through the momentum index) set to the on-shell momentum, i.e.~with $\vert \boldsymbol k^\star \vert = p^\star$. The different terms in brackets lead to terms that are sums of smooth summands and can be shuffled into the remainder term. After summing over the resulting geometric series, we obtain the following for the correlator:
\begin{align}
C_L( P) = A(P)\, {i} \left[ S^{-1}(P; L) + u \Bar{\mathcal{K}}(P) u^\dagger + g^2 T(P) \right]^{-1} A^\dagger(P) + C^{(ii)}_\infty( P) \, .
\end{align}
Using the same arguments as in section \ref{sec:luscher}, we can derive the quantization condition:
\begin{equation}
\det \left[ S^{-1}(E_n(\boldsymbol P, L), \boldsymbol P) + u \Bar{\mathcal{K}}(E_n(\boldsymbol P, L), \boldsymbol P) u^\dagger + g^2 T(E_n(\boldsymbol P, L), \boldsymbol P) \right] = 0 \,,
\end{equation}
at all finite-volume energy levels $E_n(\boldsymbol P, L)$. Given that $S^{-1}(E_n(\boldsymbol P, L), \boldsymbol P) $ and $T(E_n(\boldsymbol P, L), \boldsymbol P)$ can be calculated numerically, one can use the knowledge of the finite-volume spectrum to obtain $\Bar{\mathcal{K}}(E_n(\boldsymbol P, L), \boldsymbol P)$ as well as the coupling $g$. This object can then be linked back to the two-to-two scattering amplitude via integral equations, in a similar vein to the procedure used for the three-particle scattering formalism of refs.~\cite{Hansen:2014eka,Hansen:2015zga,Briceno:2018mlh}. We leave further discussion to the upcoming paper.

\section{Summary \& Outlook}
\label{sec:conclusion}

In this proceedings, we have described our progress in addressing issues arising in the L\"{u}scher finite-volume scattering formalism \cite{Luscher:1986pf} and extensions \cite{\justExtensions} in the case of sub-threshold finite-volume energies appearing on the $t$-channel cut. This work is motivated by recent lattice calculations in baryon-baryon systems that have observed such energy levels~\cite{Green:2021qol}.

To present the extension we first reviewed the standard derivation, following the method of Kim, Sachrajda, and Sharpe \cite{Kim:2005gf} for the case on non-identical spin-zero particles. We then identified the step in the derivation that fails to correctly account for the sub-threshold cut and provided a modification to address the issue. Our main result is an adapted quantization condition that applies above and below elastic threshold, including on the cut associated with single-meson exchange, though not on lower cuts arising from the exchange of multiple mesons.

As we have in mind applications to baryon-baryon scattering, the next step, currently ongoing, is generalizing the derivation to particles with arbitrary intrinsic spin. Once the theoretical work is concluded, future directions include numerical tests on mock data (e.g.~in the spirit of refs.~\cite{\toystudies}) and eventually applications to lattice QCD baryon-baryon data.

\section*{Acknowledgements}

The authors thank
Ra{\'ul} Brice{\~n}o,
John Bulava,
Evgeny Epelbaum,
Drew Hanlon,
Arkaitz Rodas,
Fernando Romero-López,
Maxim Mai,
Steve Sharpe, and
Hartmut Wittig
for useful discussions, including those in the context of the \emph{Bethe Forum on Multihadron Dynamics in a Box} that took place at the Bethe Center for Theoretical Physics in
Bonn, Germany.
M.T.H.~is supported by UKRI Future Leader Fellowship MR/T019956/1, and both M.T.H and A.B.R. are partly supported by UK STFC grant ST/P000630/1.

\bibliographystyle{JHEP}
\bibliography{refs}

\end{document}